# Vector Angular Spectrum Model for light travelling in scattering media

*Kaige Liu, Hengkang Zhang, Zeqi Liu, Bin Zhang, Xing Fu\*, Qiang Yuan\*, and Qiang Liu\**

K. Liu, Z. Liu, X. Fu, Q. Liu

Key Laboratory of Photonics Control Technology (Tsinghua University), Ministry of Education, Beijing 100084, China

E-mail: fuxing@mail.tsinghua.edu.cn; qiangliu@ mail.tsinghua.edu.cn

K. Liu, Z. Liu, X. Fu, Q. Liu

State Key Laboratory of Precision Measurement Technology and Instruments, Department of Precision Instrument, Tsinghua University, Beijing 100084, China

K. Liu, Q. Yuan

Research Center of Laser Fusion, China Academy of Engineering Physics, Mianyang 621900, China

H. Zhang

Beijing Institute of Control Engineering, Beijing 100190, China

B. Zhang

Beijing Institute of Electronic System Engineering, Beijing 100854, China

Strongly scattering media disrupt both the wavefront distribution and the polarization state of the incident light field. Controlling and effectively utilizing depolarization effects are crucial for optical applications in highly scattering environments, such as imaging through dense fog. However, current simulation models have difficulty simulating the evolution of vector light fields within scattering media, posing challenges for studying vector light fields in strongly scattering environments. Here, we propose the Vector Angular Spectrum (VAS) model for simulating the propagation of vector light fields within scattering media. By introducing the angular spectrum distribution of vector light scattering and polarization conversion mechanisms, this model can simulate the depolarization effects of vector light propagating through strongly scattering media. The VAS model has also been used to investigate the focusing of vector scattered light through scattering media. Furthermore, the simulation results of the model have been validated through experiments. The proposed VAS model is expected to play a role in the theoretical research of vector scattered light and optical applications in strongly scattering environments.

## 1. Introduction

Scattering media, such as turbid water, biological tissues, multimode optical fibers, etc., have the ability to disturb the propagation process of the light field[1, 2]. The wavefront of the light travelling through them is destroyed, thus losing the original propagation path, and the coherent light field will form the speckle pattern. The amplitude and phase information of the light wave is disrupted[3]. In addition, the highly scattering media will destroy the polarization characteristics of the light field, and the input polarized light will be less polarized as it propagates deeper[4, 5]. Thanks to the in-depth study of scattering phenomena, researchers have realized focusing[6, 7] and imaging[8, 9] in scattering environment, which greatly facilitating applications in biomedicine[10-12], optical tweezers[13, 14], optical communication[15, 16] and many other research areas[17-19]. The addition of polarization information provides more degrees of freedom, further broadening the scope of application of the scattering media, such as imaging through the fog[20], polarization control of multimode optical fibers[7] and other aspects of the application[21].

In the research of these applications, a reliable simulation model can greatly assist theoretical studies and quantitative analysis. Currently, there are two main simulation models, including the Monte Carlo (MC) model[22-25] and the Angular Spectrum (AS) model[26-29]. Among them, the MC model simulates the random propagation process of a large number of photons and accumulates the results, achieving high simulation accuracy and having the ability to simulate the transmission of vector light. However, an accurate simulation often requires a large number of input photons, making the simulation process time-consuming. Moreover, the MC model requires the setting of receivers at the target positions, which means that photons stop propagating once they reach the observation points, making it difficult to observe the complete transmission process within the scattering media. In contrast, the AS model layers the scattering media and replaces scattering events with phase masks, directly simulating the evolution process of the complex optical field transmission within the scattering media. The AS model has a faster computation speed and is more advantageous when studying the scattering transmission process. Furthermore, its improved model full-polarization angular spectrum model (*fp*ASM) has further taken polarization into account by introducing an additional effect of scattering angle in the random phase mask, thus enabling the AS model to simulate the transmission of vector light[29]. Although *fp*ASM can effectively simulate the scattering phenomenon after fully depolarization, it is difficult for the model to accurately quantify the evolution process of the light field from linearly polarized input to fully depolarization due to the lack of consideration of the physical mechanism of vector light scattering in the scattering medium during polarization coupling. The issue of MC model's inability to simulate the internal evolution of polarized light remains unresolved.

In this article, we propose a vector angular spectrum (VAS) model based on the vector optical radiation theory for simulating the transport of vector light inside a scattering medium. Using the VAS model, we can simulate the process of transmission, scattering, and polarization transformation of a polarized light field inside a scattering medium. By comparing the simulation data with experimental data, we verify the validity of the VAS model. Using VAS model, we further simulate vector light focusing through scattering medium with the vector transmission matrix, and study the controllable range of the polarization state of the focus and verify it experimentally.

## 2. Principle
### 2.1. Phase mask calculation

The propagation of the VAS model comprises three computational stages, as illustrated in **Figure.1**. The scattering effects within the scattering medium are decoupled into three processes: phase delay, polarization coupling and free-space transmission. The scattering medium is divided into N layers. The light field undergoes these three stages as it passes through each layer. Firstly, a random phase mask is superimposed on the input optical field to simulate the change in the direction of light transmission after passing through a scattering layer. This scattering effect is abstracted as a phase delay, which can be physically interpreted as the difference in average refractive index at different spatial positions. For the entire scattering medium, there is an average refractive index $n_0$, while the random variations in local refractive index at different spatial positions can be approximated by a Gaussian distribution. The phase delay caused by this Gaussian distribution has two main effects: the magnitude of the standard deviation determines the magnitude of phase delay fluctuations across the entire plane, which is associated with the attenuation of ballistic light during the scattering process, and further determines the scattering mean free path; the relative phase delay between adjacent elements on the plane determines the distribution of scattering angles, and thus the magnitude of the forward scattering coefficient *g* of the scattering medium.

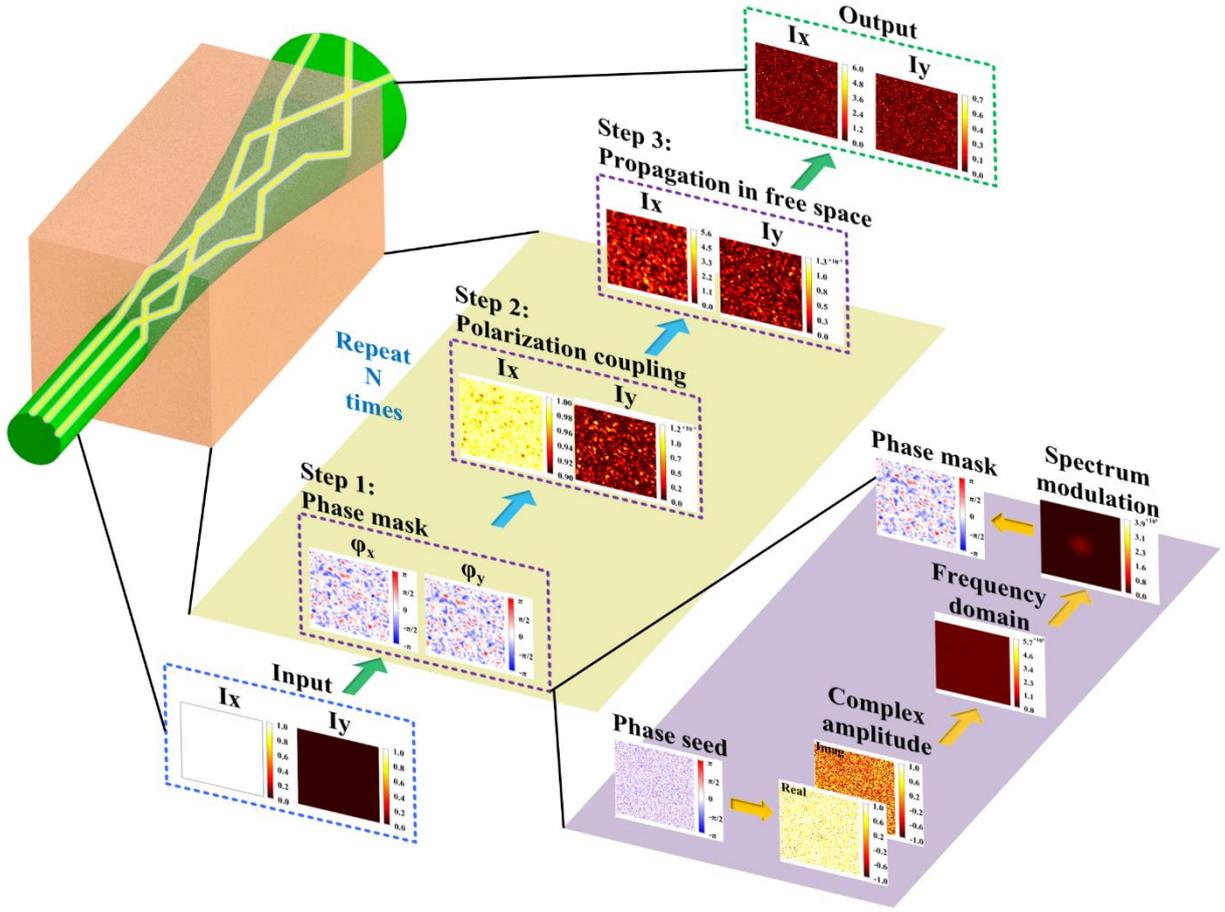

**Figure 1.** Schematic diagram of the VAS model.

Firstly, random numbers conforming to a Gaussian distribution $N(0, \sigma_p^2)$ are generated on a simulation grid as random phases, with each grid cell independently generated. The standard deviation $\sigma_p$ of this Gaussian distribution determines the scattering mean free path $l_s$ of the simulated scattering medium. The magnitude of $\sigma_p$ can be calculated based on the physical meaning of $l_s$. Since $l_s$ characterizes the strength of scattering, manifesting in the attenuation rate of ballistic light and can be expressed as[30]:

$$I_D = I_0 \exp(-l/l_s), \tag{1}$$

where $I_D$ is the intensity of the ballistic light, $I_0$ is the intensity of the incident light and $l$ is the depth of light penetration. The propagation of the optical field is represented in the form of electric fields, making it convenient to obtain frequency-domain information through Fourier transform. Ballistic light represents photons without scattering, and when the input optical field is a plane wave, the component of ballistic light can be obtained from the value at $k_x = k_y = 0$ in the frequency domain.

Furthermore, considering the phase distribution pattern after superimposing different layers of phases, each phase mask follows the same Gaussian distribution $N(0, \sigma_p^2)$. The free-space transmission process does not affect the phase distribution pattern but only changes the planar distribution position of the phase. Therefore, the phase distribution after $n$ layers of scattering remains Gaussian, with the standard deviation becoming $n\sigma_p$. If the layer spacing of the simulation model is $d$, then the depth of light penetration is $l = nd$. Next, we consider the relationship between the phase distribution at a penetration depth of $l_s$ and the proportion of ballistic light. After Fourier transforming the complex amplitude, the value at the center of the frequency domain represents the mean value in the spatial domain. Hence, the proportion of ballistic light can be expressed as:

$$I_D = \left| \langle E(x,y) \rangle \right|^2 = \langle A \cos \varphi \rangle^2 + \langle A \sin \varphi \rangle^2, \quad (2)$$

where $A$ and $\varphi$ are the amplitude and phase of the light field, respectively. As previously mentioned, the superposition of the $n$-layer model is equivalent to using a single-layer phase mask with an $n$-times standard deviation phase. Therefore, to simplify the calculation, we consider the superposition of a single layer here. When the input is a plane wave, $A$ is constantly 1. Since $\varphi \sim N(0, \sigma_p^2)$, we have:

$$\begin{aligned} \langle A \sin \varphi \rangle &= 0 \\ \langle A \cos \varphi \rangle &= \frac{1}{\sqrt{2\pi}\sigma_p} \int_{-\infty}^{\infty} \cos \varphi \exp\left(\frac{-\varphi^2}{2\sigma_p^2}\right) d\varphi = \exp\left(-\sigma_p^2/2\right) \end{aligned}. \quad (3)$$

Then we can get the value of the standard deviation:

$$\sigma_p = \sqrt{d/l_s}. \quad (4)$$

The standard deviation determines the magnitude of random phase fluctuations, which in turn defines the scattering mean free path. On the other hand, the rate of spatial variation in the phase mask determines the scattering angle. To satisfy the scattering angle distribution in a realistic scattering model, it is necessary to constrain the spatial distribution of random phases while maintaining the overall statistical distribution. As photons traverse a scattering medium, each scattering event occurs according to a certain probability model, and the overall scattering phenomenon is the superposition of countless such events. The scattering patterns at specific depths can be calculated using the radiative transfer equation. The scattering field component can be expressed as[31]:

$$I(\mathbf{r}, \hat{s}) = I_{ri}(\mathbf{r}, \hat{s}) + I_d(\mathbf{r}, \hat{s}), \quad (5)$$

where $I_{ri}$ represents the reduced incident intensity and $I_d$ represents the diffuse intensity, $\mathbf{r}$ is the position vector and $\hat{s}$ is the direction vector. When we layer the scattering medium, the spacing between each layer is small, usually less than the scattering mean free path, making it suitable for the sparse first-order approximation[31]. The diffuse intensities for forward and backward scattering can be calculated with the following expression:

$$I_{d+}(\tau,\mu,\mu_0,\phi) = \int_0^\tau \exp\left[-\frac{(\tau-\tau_1)}{\mu} - \frac{\tau_1}{\mu_0}\right] \cdot \frac{p(\mu,\phi;\mu_0,\phi_0)}{4\pi} F_0 \frac{d\tau_1}{\mu}$$
$$= \frac{p(\mu,\phi;\mu_0,\phi_0)}{4\pi} \frac{\exp(-\tau/\mu_0) - \exp(-\tau/\mu)}{(\mu_0-\mu)} \mu_0 F_0, \tag{6}$$

$$I_{d-}(\tau,\mu,\mu_0,\phi) = \int_\tau^{\tau_0} \exp\left[-\frac{(\tau-\tau_1)}{\mu} - \frac{\tau_1}{\mu_0}\right] \cdot \frac{p(\mu,\phi;\mu_0,\phi_0)}{4\pi} F_0 \frac{d\tau_1}{-\mu}$$
$$= \frac{p(\mu,\phi;\mu_0,\phi_0)}{4\pi} \frac{\exp(-\tau/\mu_0) - \exp\left[-\tau_0/\mu_0 + (\tau_0-\tau)/\mu\right]}{(\mu_0-\mu)} \mu_0 F_0, \tag{7}$$

where $I_{d+}$ and $I_{d-}$ are the forward and backward diffuse intensities, respectively. $\tau = l/l_s$ is the optical depth, $\mu = \cos(\theta)$, $\mu_0 = \cos(\theta_0)$, $\theta_0$ and $\theta$ are the scattering angle before and after the scattering event, $\phi$ is the azimuthal angle and $F_0$ is the initial incident radiation flux. The scattering phase function is given by[23]:

$$P(\theta,\phi) = s_{11}(\theta)I_0 + s_{12}(\theta)\left[Q_0\cos(2\phi) + U_0\sin(2\phi)\right], \tag{8}$$

where $s_{11}$ and $s_{12}$ are elements of scattering matrix of a single scattering event, $S_0 = [I_0, Q_0, U_0, V_0]$ is the stokes vector of the incident light.

In angular spectrum propagation, we only consider the forward-propagating components. Therefore, the proportion of components with different scattering angles and azimuth angles after scattering can be calculated using Equation. 6. For a phase mask, it is equivalent to a layer of solidified scattering medium that adds different phase delays to different spatial positions. Consequently, the incident light in the above calculations is taken as normally incident, i.e., $\mu_0 = 1$. Polarized light fields can be decomposed into horizontal and vertical polarization directions. Taking the example of horizontal polarization, we have $I_0 = Q_0 = 1$ and $U_0 = 0$. The values of $s_{11}$ and $s_{12}$ are determined by the scattering properties of the scattering medium. For the most commonly encountered spherical particles in experiments, their scattering matrix can be solved using Mie scattering theory[32]. When a scattering medium is given for simulation, the spectral distribution of the scattered light field after a parallel beam of normally incident light passes through the phase mask can be calculated using Equation. 6 and Equation. 8. For the phase mask generated using Gaussian random numbers mentioned earlier, we can calculate its scattered light field spectral distribution, as shown in **Figure. 2**a1-a3, presenting a nearly uniform distribution. By superimposing the calculated spectral intensity distributions, we obtain the distribution shown in Figure. 2b1-b3, which satisfies the outgoing spectral distribution for scattering angles and azimuth angles. After converting this result to a complex amplitude distribution using the inverse Fourier transform, we extract the phase distribution as the final scattering phase mask.

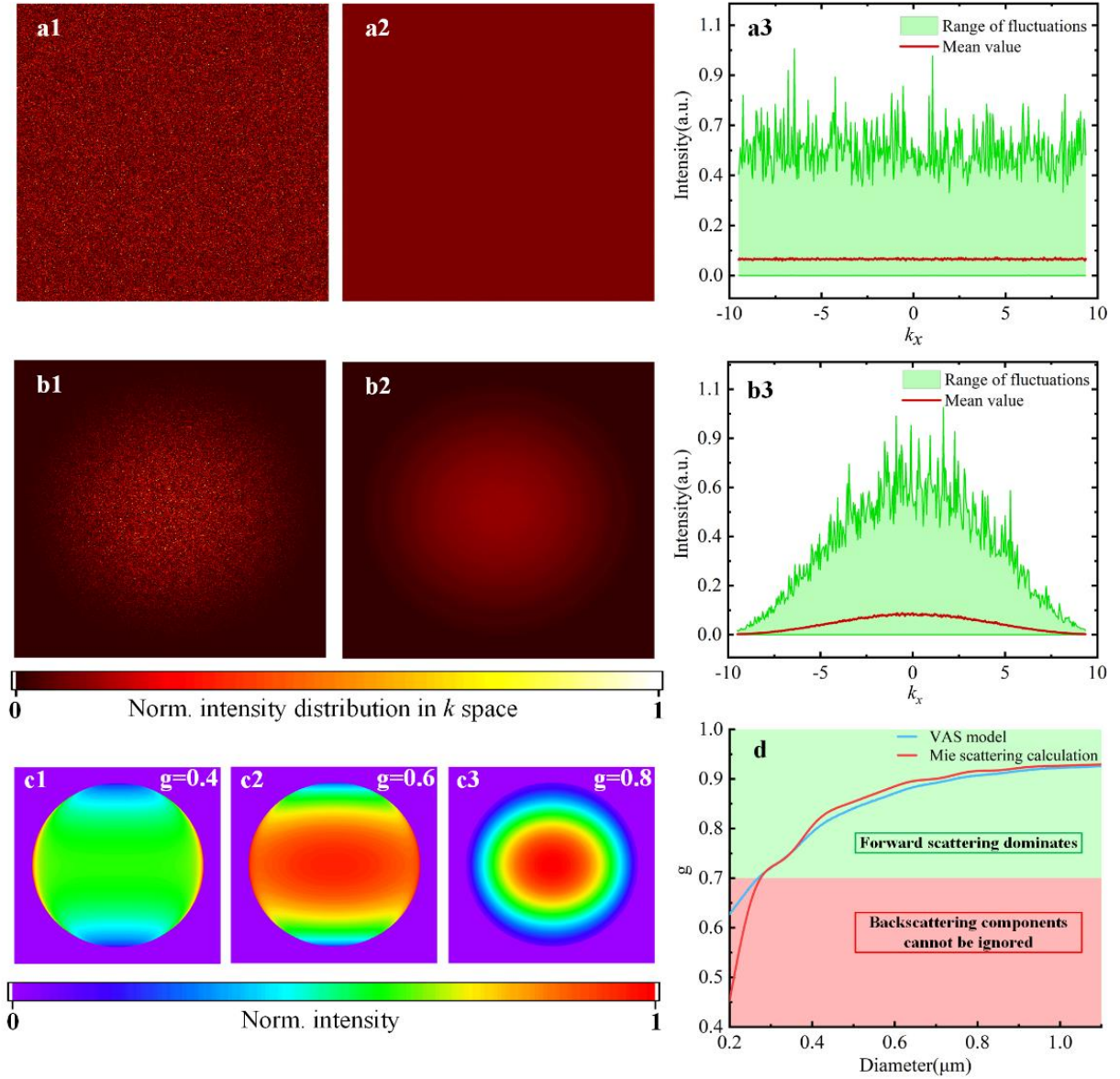

**Figure 2.** Calculation of random phase mask. a1-a3) The frequency distribution of Gaussian random phases with single calculation, averaged by 100 times calculations and for $k_x = 0$, respectively; b1-b3) the frequency distribution of altered phase mask with single calculation, averaged by 100 times calculations and for $k_x = 0$, respectively; c1-c3) spectral intensity distributions for $g = 0.4, 0.6, 0.8$; d) $g$ value calculated by VAS model and Mie theory with different diameters of the particles.

For the scattering of polarized light, the distribution of azimuth angles is not uniform over 0 to $2\pi$. As the scattering angle changes, the probability of azimuth angle values also varies. Consequently, the scattering of polarized light exhibits a significant dependence on the incident polarization state. By adjusting the forward scattering coefficient $g = <\cos\theta>$, we can obtain the scattering angle distribution under different conditions, as shown in Figure. 2c1-c3. The lower the $g$ value, the greater the difference in scattering between the co-polarized and cross-polarized directions.

To validate the accuracy of the scattering phase mask obtained from the VAS model in terms of scattering angles, we calculated the forward scattering coefficient $g$ for spherical particle solutions with different particle diameters as the scattering medium and compared it with the results obtained using Mie theory, as shown in Figure. 2d. In the VAS model, we performed a Fourier transform on the complex amplitude of scattering field at a penetration depth of one scattering mean free path to obtain the frequency domain distribution. With $k_x = \sin\theta\cdot\cos\varphi$ and $k_y = \sin\theta\cdot\sin\varphi$, we can obtain the proportions of different scattering angles and subsequently calculate $g$. The results from the VAS model closely align with those from Mie theory in the range of $g > 0.7$. However, in the range of $g < 0.7$, the VAS model overestimates the results compared to Mie theory. This is because the VAS model's calculations only consider forward-propagating components, and when the $g$ value is small, the contribution of backward components becomes non-negligible. When calculating $g$, the backward components correspond to negative values of $\cos\theta$, so ignoring this part leads to an overestimation of $g$. Common scattering media encountered in applications, such as biological tissues, often have high $g$ values, typically $g > 0.9$, with forward scattering dominating. Therefore, in most application scenarios, ignoring backward scattering does not introduce significant errors.

## 2.2. Polarization transformation

After applying the phase mask, the model completes the simulation of the scattering process. However, for the scattering of vector light, the polarization state also changes. The depolarization effect during the transmission of polarized light within the scattering medium is a complex nonlinear effect. And there have been many theoretical studies focusing on this process and quantitative analysis[33-35]. Here, we adopt a model that approximates the change in polarization angle with the penetration depth. After a certain penetration depth, the rotation angle of the polarization direction of the polarized light approximately satisfies a Gaussian distribution[34]:

$$p(\alpha) = \frac{\exp\left(-\dfrac{\alpha^2}{2\chi\tau}\right)}{\sqrt{\chi\tau}\cdot\sqrt{\pi/2}}, \tag{9}$$

where $\alpha$ represents the polarization angle deviation after penetrating a depth of $\tau$, $\tau$ is the optical depth defined as the thickness of the scattering medium divided by the scattering mean free path, and $\chi$ characterizes the speed of polarization conversion and is only related to the physical properties of the scattering medium. Then, the degree of polarization (DOP) can be calculated with:

$$P_L = \frac{I_{par} - I_{per}}{I_{par} + I_{per}} = \exp\left(-\frac{\chi\tau}{2}\right), \tag{10}$$

where $I_{par}$ and $I_{per}$ are intensities of light parallel and perpendicular to the incident light, respectively. Tracking the scattering process of a single photon, each scattering event corresponds to a specific scattering phase function, and multiple scatterings represent the statistical superposition of the probability distribution of the phase function. For a scattering medium, the scattering mean free path represents the average distance a single photon travels before experiencing a scattering event. Therefore, once the scattering mean free path and the thickness of the scattering medium are determined, the expected state of the photon upon final exit can be calculated. Here, we adopt the results from ref[36] and can calculate the value of $\chi$ using the following equation:

$$\delta = \left(\int \frac{s_{11}(\theta) - s_{33}(\theta)}{2} d\mu\right) \Big/ \left(\int s_{11}(\theta) d\mu\right), \tag{11}$$

$$\chi = \mu_s \sqrt{(2\delta + 1 - g) \cdot (1 - g)}, \tag{12}$$

where $s_{33}$ is element of the scattering matrix of a single scattering event, $\mu_s$ is scattering coefficient.

The calculation of depolarization between layers requires distinguishing the conversion between scattered light and ballistic light, as their conversion ratios differ. First, considering the conversion of ballistic light, from the $n$ th layer to the $(n+1)$ th layer, the light travels from $z = nd$ to $z' = (n+1)d$ along the propagation direction. A certain proportion $\left(\exp(-nd\mu_s) - \exp(-(n+1)d\mu_s)\right)$ of the ballistic light is lost and converted into scattered light, but it has already traveled a distance of $(n+1)d$. Therefore, the depolarization conversion ratio for this portion of light is given by:

$$c_{par1} = \frac{1}{2}\left(1 + \exp\left(-\frac{\chi(n+1)d}{2}\right)\right), \quad c_{per1} = \frac{1}{2}\left(1 - \exp\left(-\frac{\chi(n+1)d}{2}\right)\right). \tag{13}$$

The magnitude of the polarized light resulting from this conversion is:

$$I_{par1} = \frac{1}{2}\left(1+\exp\left(-\frac{\chi(n+1)d}{2}\right)\right)\left(\exp(-nd\mu_s)-\exp(-(n+1)d\mu_s)\right),$$

$$I_{per1} = \frac{1}{2}\left(1-\exp\left(-\frac{\chi(n+1)d}{2}\right)\right)\left(\exp(-nd\mu_s)-\exp(-(n+1)d\mu_s)\right).$$

(14)

The scattered light accumulated in previous layers also undergoes polarization conversion. Since the depolarization effects have already been accounted for in the previous layers, the conversion ratio between two layers is:

$$c_{par2} = \frac{1}{2}\left(1+\exp\left(-\frac{\chi d}{2}\right)\right), c_{per2} = \frac{1}{2}\left(1-\exp\left(-\frac{\chi d}{2}\right)\right). \quad (15)$$

And the amount of polarized scattered light accumulated in the $n$th layer is:

$$I_{par2} = \frac{1}{2}\left(1+\exp\left(-\frac{\chi nd}{2}\right)\right)(1-\exp(-nd\mu_s)),$$

$$I_{per2} = \frac{1}{2}\left(1-\exp\left(-\frac{\chi nd}{2}\right)\right)(1-\exp(-nd\mu_s)).$$

(16)

After polarization conversion, the polarized light component generated by this portion of light is:

$$I_{par3} = I_{par2} \cdot c_{par2} + I_{per2} \cdot c_{per2} = \frac{1}{2}(1-\exp(-nd\mu_s))\left(1+\exp\left(-\frac{\chi(n+1)d}{2}\right)\right). \quad (17)$$

Finally, combining these components, we obtain the total co-polarized light component:

$$I_{par} = I_{par1} + I_{par3} = \frac{1}{2}(1-\exp(-(n+1)d\mu_s))\left(1+\exp\left(-\frac{\chi(n+1)d}{2}\right)\right). \quad (18)$$

At this point, the proportion of co-polarized light in the polarized light is $\frac{1}{2}\left(1+\exp\left(-\frac{\chi(n+1)d}{2}\right)\right)$, which is consistent with the predicted ratio from the model.

### 2.3. Transmission process

After polarization conversion, the process of vector scattering is completed, followed by a free transmission process to simulate the propagation and coherent superposition of scattered light. Firstly, the attenuation effect of energy loss during backward propagation on the light field intensity should be considered. By integrating over the backward angle using Equation. 7, the backward scattered light component of the light field passing through a layer of scattering medium from any incident direction can be calculated as:

$$F_-(\mu_0) = \int_0^{2\pi}\int_{-1}^{0} I_{d-}(\tau,\mu,\mu_0,\phi)\mu d\mu d\phi. \quad (19)$$

After the aforementioned two stages, the Fourier transform of the obtained light field can be performed to determine the proportion of different spectral components. For each spectral component, its energy retention rate after penetrating a layer is calculated and multiplied to the intensity of that component. After completing the attenuation of the intensity of all frequency components, an inverse Fourier transform is applied to convert it back to the spatial domain.

Finally, the free transmission process is performed. For the angular spectrum propagation of vector light fields, it can be calculated using the following equation[37]:

$$E_x(x,y,z) = \iint_\infty \mathbf{A}_x(p,q) e^{ik(px+qy+mz)} dpdq,$$

$$E_y(x,y,z) = \iint_\infty \mathbf{A}_y(p,q) e^{ik(px+qy+mz)} dpdq, \quad (20)$$

$$E_z(x,y,z) = \iint_\infty \left[\frac{p}{m}\mathbf{A}_x(p,q) + \frac{q}{m}\mathbf{A}_y(p,q)\right] e^{ik(px+qy+mz)} dpdq,$$

where $\mathbf{A}_x$ and $\mathbf{A}_y$ are angular spectrum components of $x$ and $y$ direction, respectively. $p$, $q$, $m$ are directional cosines of $x$, $y$ and $z$ directions, respectively. However, in the model, the electric field component in the $z$-direction is much smaller than the $xy$-components and does not participate in the propagation, so it is not considered in the calculation. After these three steps, the vector angular spectrum transmission for a single layer is completed. The output light field then enters the next layer, repeating these three steps until all layers are calculated, yielding the final output light field transmitted through the scattering medium.

## 3. Results

### 3.1. Light propagation

We first simulated the beam propagation process, using 2000 × 2000 input pixels with a pixel size of 1 μm. The input light field was a rectangular $x$-polarized beam centered on a 400 × 400 area, with a wavelength of 532 nm. The scattering medium was set as an aqueous solution of polystyrene microspheres with a particle diameter of 500 nm. The refractive index of the scattering particles was $n_1 = 1.59$, and the refractive index of the medium was $n_2 = 1.33$, with a dilution factor of 0.3%. Based on these parameters, the scattering mean free path of the medium was calculated to be $l_s = 100$ μm, with a forward scattering coefficient of $g = 0.854$. The total thickness of the scattering medium was set to 2 mm, divided into 40 layers with a layer spacing of 50 μm, which is less than one scattering mean free path. Using the VAS model, we could quickly obtain the light field distribution within the 40 layers of the scattering medium. Partial intensity distribution maps of some layers are shown in **Figure. 3**b. At one scattering mean free path, photons undergo an average of one scattering event, exhibiting initial speckle characteristics, with minimal depolarization effects. At one transport mean free path, $l' = l_s / (1-$ *g*), ballistic light is completely overwhelmed by scattered light, and the light field exhibits irregular speckle characteristics. Upon reaching a penetration depth of 20 times the scattering mean free path, depolarization effects become significant, with the intensity in the *y*-direction reaching comparable levels to that in the *x*-direction.

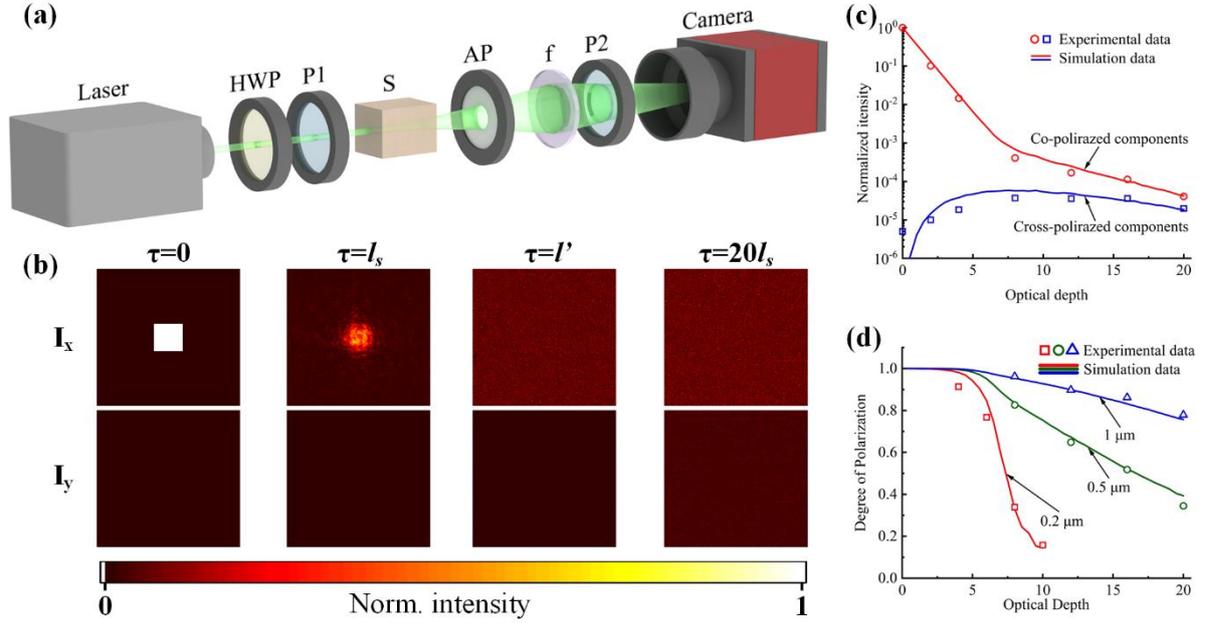

**Figure 3.** Light transmission in turbid medium. a) Experimental setup, HWP: half-wave plate, P1, P2: polarizer, S: scattering medium, AP: aperture, f: tube lens; b) evolution of the light field distribution with increasing penetration depth; (c) intensity transmission as a function of the optical depth; d) degree of polarization as a function of the optical depth with different diameters of the particles (0.2μm, 0.5μm, 1μm).

To quantitatively validate the effectiveness of the VAS model, we experimentally tested the changes in light field intensity and polarization through scattering media of different depths. The experimental setup is shown in Figure. 3a. A 532 nm laser beam (home-made, solid-state, continuous-wave, 6 W) was first modulated in intensity and polarization by a combination of a half-wave plate and polarizer, before being incident on a cuvette containing the scattering sample. The transmitted light passed through a small aperture to filter out most of the scattered light except for that exiting near the center position. The light field was then collected by a tube lens ($f = 180$ mm) and focused onto a complementary metal oxide semiconductor (COMS) camera (PCO, edge 4.2 bi, pixel size: 6.5 μm × 6.5 μm). To calculate the transmission, the light field intensity was calculated only at the center $10 \times 10$ pixels of the camera. In the experiment, three different polystyrene microsphere diameters (200 nm, 500 nm, 1000 nm) were used, and the optical thickness of the sample was controlled by using different thicknesses of cuvettes. The experimentally measured intensity transmittance for the 500 nm diameter sample is shown in Figure. 3c, and the results of the DOP decrease with penetration depth for the three samples are shown in Figure. 3d. In the figures, the experimental results are represented by hollow symbols, while the simulation results are averaged from 100 simulations and represented by solid lines. The above results show that the simulation results from the VAS model closely match the experimental findings.

**3.2. Focusing of vector light through a scattering medium**

The polarization changes caused by penetrating strongly scattering media can be controlled using wavefront shaping techniques. Through wavefront shaping, light focusing can be achieved through scattering media, and by incorporating the freedom of polarization, further control over the polarization state of the light field can be achieved. Currently, there have been many reports on the research of vector scattered light control, including the regulation of polarized light proportions[6] and the generation of focal points with arbitrary polarization states[7, 18]. The properties of scattering media, including their optical thickness, forward scattering coefficient, etc., can have a significant impact on the application of these technologies. Utilizing the VAS model, we can conduct quantitative analysis on different parameters to study the advantages and limitations of these technologies. Using the VAS model, we simulated the process of focusing vector scattered light using the transmission matrix method. The simulation employed 2000 × 2000 input pixels with a pixel size of 1μm, a wavelength of 532 nm, and an input light field that was *x*-polarized and occupied the central 512 × 512 area. The transmission matrix was measured using a 1024-order Hadamard basis. And each Hadamard vector was shaped into a 32 × 32 plane distribution, with each input module occupying 16 × 16 input pixels. The scattering particles in the medium had a diameter of 500 nm, a refractive index of $n_1$ = 1.59, a base refractive index of $n_2$ = 1.41, and a dilution factor of 0.31%. Based on these parameters, the scattering mean free path was calculated to be $l_s$ = 200 μm, with a forward scattering coefficient $g$ = 0.8734. The layer spacing of the scattering medium was fixed at 100 μm, with layer numbers of 10, 20, 30, 40, and 50, corresponding to optical depths of 5, 10, 15, 20, and 25. The imaging plane was set at a distance of 1 mm from the rear surface of the scattering medium. The experimental setup used for comparison is shown in **Figure. 4**a. A 532 nm wavelength laser (home-made, solid-state) with a maximum output power of 6 W was used as the light source. The emitted laser beam first passed through a beam expander to increase the beam diameter, fully utilizing the control units of the spatial light modulator (SLM). Then, a combination of a half-wave plate and polarizer was used to modulate the light intensity and polarization state. After the beam was irradiated onto the SLM and modulated, the +1 diffraction order was selected with a 4-f system ($f_1$ = 600 nm, $f_2$ = 300 nm)[38]. After passing through the scattering medium, the transmitted light field was magnified using an objective lens (TU Plan ELWD 50×, Nikon, NA = 0.6) and finally imaged onto a CMOS camera (PCO, edge 4.2 bi, pixel size: 6.5 μm × 6.5 μm) through a tube lens ($f$ = 180 mm). The scattering media used in the experiment were made by embedding 500 nm diameter polystyrene microspheres into a cured polydimethylsiloxane (PDMS) polymer (SYLGARD 184), with thicknesses of 1, 2, 3, 4, and 5 mm[39].

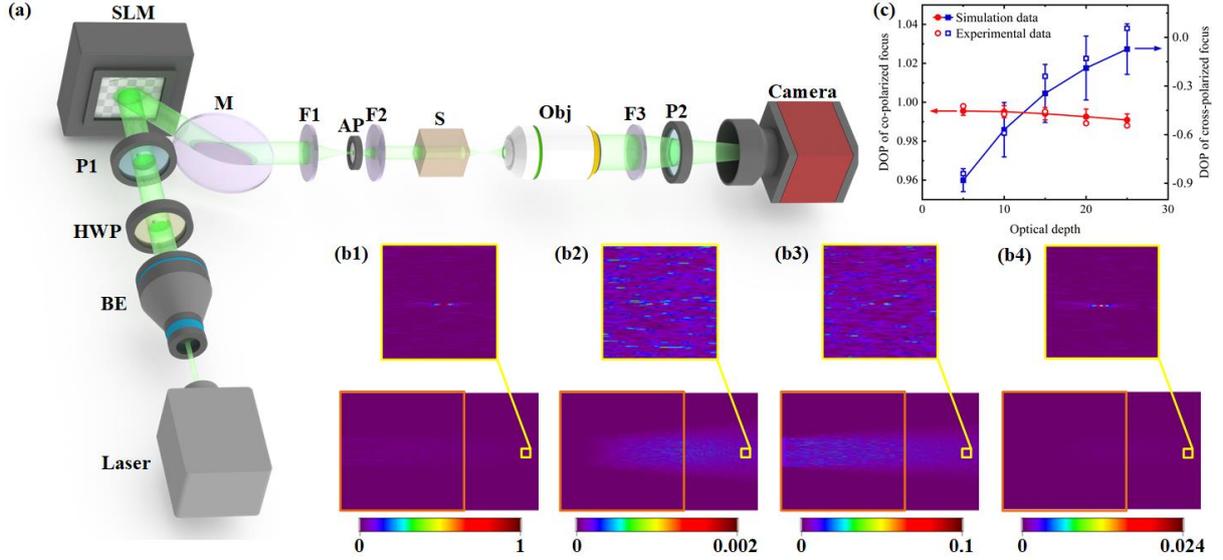

**Figure 4.** Focusing of vector light through the scattering medium. a) Experimental setup, BE: beam expander, HWP: half-wave plate, P1, P2: polarizer, SLM: spatial light modulator, M: mirror, F1, F2, F3: lens, AP: aperture, S: scattering medium, Obj: objective; b1-b2) focus formed with Tx; b3-b4) focus formed with Ty; c) DOP of co-polarized and cross-polarized focus.

Figure. 4b1-b4 demonstrate the evolution process of vector scattered light focusing through a scattering medium with an optical depth of 10 using the VAS model simulation. After inputting the light field with *x*-polarization, the output light field used for measuring the transmission matrix can be obtained from either the co-polarized (*x*-direction) or cross-polarized (*y*-direction) output distribution. The transmission matrix calculated from the co-polarized output is denoted as $T_x$, while the transmission matrix calculated from the cross-polarized output is denoted as $T_y$. When using $T_x$ for focusing, a sharp focal point is formed at the target position in the co-polarized direction, as shown in Figure. 4b1, while there is also some intensity at the same position in the cross-polarized direction, as shown in Figure. 4b2. Therefore, the polarization state formed at this time is not the desired *x*-direction linear polarization. The intensity ratio in the *x-y* direction of the focal point is 526.31. When using $T_y$ for focusing, focus is achieved at the target position in the cross-polarized direction, as shown

in Figure. 4b4, but due to the presence of a large amount of light without polarization transformation forming speckles in the *x*-polarized direction, as shown in Figure. 4b3, the intensity ratio of the focal point in the *y*-*x* direction is only 0.24. This indicates that when the polarization state of the target focus is perpendicular to the input light polarization state, the proportion of the target polarization state in the polarization components of the focal point is even lower than that of the vertical component. Furthermore, let's illustrate the impact of this phenomenon by using an example of polarization focusing based on the transmission matrix. To achieve a polarized focus in the *x*-direction, when the input is equally proportioned in the *x*-*y* polarization directions, the *x*-direction component of the focal point is 0.5 × 526.31/(526.31+1)+0.5×0.24/(0.24+1) = 0.5958, while the *y*-direction component is 0.4042. At this point, the actual polarization degree of the generated focal spot is only 0.1916, rather than the expected linearly polarized light. Therefore, the direct method of focusing often finds it difficult to achieve complete control over the polarization state, and when obtaining a light focus with the desired polarization direction, it is difficult to achieve a high level of polarization purity[6]. However, based on the experiments and simulations in Section 3.1, we can see that as the penetration depth increases, the depolarization effect becomes more pronounced. This suggests that the proportion of the cross-polarized light component increases, and there will be less speckle intensity from the co-polarized light during cross-polarized focusing. Therefore, we further conducted experimental and simulation studies on vector scattered light focusing for the aforementioned five scattering media and measured the DOP of the focal points achieved using $T_x$ and $T_y$ for focusing. The experimental results align well with the simulation results as shown in Figure. 4c, and the negative DOP indicates that the proportion of the target polarization component is less than that of its perpendicular polarization direction. As the penetration depth increases, the DOP of the co-polarized focusing hardly changes and remains at a high level (greater than 0.98). The polarization degree of the cross-polarized focusing increases from nearly -1 to approximately 0. Therefore, for direct polarization focusing methods, a larger optical depth of the scattering medium is conducive to improving the polarization purity of the focal spot.

## 4. Conclusion

In this article, we propose the VAS model for simulating the propagation of vector light within scattering media. The VAS model enables the simultaneous acquisition of polarized light field distributions across tens of planes, allowing for the investigation of the process of vector light propagation within scattering media. Utilizing the VAS model, we have studied the polarization transformation of light within scattering media, and the simulated curves of polarization degree and transmittance in two polarization directions versus penetration depth align well with the experimental results. Additionally, using the VAS model, we have also investigated the process of focusing vector light through scattering media using wavefront shaping techniques, explaining the reasons for the difficulty in obtaining a fully linearly polarized focal point. Our experimental results have validated the simulation outcomes.

**Acknowledgements**

This work was supported by the National Natural Science Foundation of China (NSFC) under Grant Number 62275137.

**Conflicts of Interest**

The authors declare no conflicts of interest.

**Data Availability Statement**

Data underlying the results presented in this paper are not publicly available at this time but may be obtained from the authors upon reasonable request.